\documentclass[twocolumn,prl,amsmath,amssymb]{revtex4}
\usepackage{graphicx}

\def\s{{\sigma}}
\def\e{{\epsilon}}
\def\k{{ {\bf k} }}

\def\q{{ {\bf q} }}

\def\w{{\omega}}

\begin{document}

\def\runtitle{
Origin of the Weak Pseudo-gap Behaviors in Na$_{0.35}$CoO$_2$
}
\def\runauthor
{Keiji {\sc Yada} and Hiroshi {\sc Kontani}}

%\draft
\title{
Origin of the Weak Pseudo-gap Behaviors in Na$_{0.35}$CoO$_2$:\\
Absence of Small Hole Pockets
}

\author{
Keiji {\sc Yada} and
Hiroshi {\sc Kontani}
}

\address{
Department of Physics, Nagoya University,
Furo-cho, Nagoya 464-8602, Japan.
}

\date{\today}

\begin{abstract}
%To elucidate the topology of the Fermi surfaces,
We analyze the {\it normal electronic states} of Na$_{0.35}$CoO$_2$
based on the effective $d$-$p$ model with full $d$-orbital freedom 
using the fluctuation-exchange (FLEX) approximation.
%Here, the topology of the Fermi surfaces
%is very sensitive to the crystalline electric
%splitting (CES) due to trigonal deformation.
They sensitively depend on the topology of the Fermi surfaces,
which changes as the crystalline electric
splitting (CES) due to the trigonal deformation.
%between $a_{1g}$-orbital and $e_g'$-orbitals, in $t_{2g}$-ones.
We succeed in reproducing the weak pseudo-gap
behaviors in the density of states (DOS) 
and in the uniform magnetic susceptibility
below 300K, assuming that six small hole-pockets
predicted by LDA band calculations are {\it absent}.
When they exist, on the contrary,
then ``anti-pseudo-gap behaviors'' should inevitably appear.
Thus, the present study strongly supports the
absence of the small hole-pockets in Na$_{0.35}$CoO$_2$,
as reported by recent ARPES measurements.
A large Fermi surface around the $\Gamma$-point would account for 
the superconductivity in water-intercalated samples.
\end{abstract}

%\pacs{}

\sloppy

\maketitle

\vspace{5mm}\noindent
[KEYWORDS: Na$_x$CoO2, FLEX approximation, \\
weak pseudo-gap, spin-fluctuations ]
\vspace{5mm}

%%%%%%%%%%%%%%%%%%%%%
% Introduction
%%%%%%%%%%%%%%%%%%%%%
Na$_{0.35}$CoO$_2 \cdot$1.3H$_2$O
is the first Co-oxide superconductor ($T_{\rm c}\approx 4.7$K)
with a triangular lattice structure
 \cite{Takada}.
Below $T_{\rm c}$, the coherent peak in $1/T_1T$ by NMR/NQR
is absent (or very tiny even if it exists)
 \cite{Sato,Tei,Ishida},
and the Knight shift is finite both for
${\bf H}\perp {\bf c}$ \cite{Sato}
and ${\bf H}\parallel {\bf c}$ \cite{Sato-single}.
Moreover, the specific heat data below $T_{\rm c}$
cannot be explained by an isotropic $s$-wave BCS theory;
one should assume an anisotropic $s$-wave gap or 
a gap with line-nodes to fit the observed data,
depending on samples \cite{Phillips}.
These experiments suggests that
the superconductivity (SC) in Na$_{0.35}$CoO$_2 \cdot$1.3H$_2$O
is not an isotropic $s$-wave superconductor.

Precise knowledge about the 
shape of the Fermi surfaces (FS's) is indispensable
in analyzing the origin of the unconventional SC.
As for Na$_{0.35}$CoO$_2$, however,
even the topology of FS's is still uncertain unfertunately.
The LDA band calculation performed
by Singh predicted that a large hole-like FS composed of 
$a_{1g}$-orbital around $\Gamma$-point, surrounded by
six small hole-pockets composed of $e_{g}'$-orbitals
 \cite{Singh}.
The latter give major contribution to the density of states 
(DOS) at the Fermi level.
The crystalline electric splitting (CES) of Co 3$d$-orbitals
is shown in Fig. 1.
Based on the Singh's FS's,
various theoretical groups have predicted 
unconventional SC's due to the Coulomb interaction
by using the RPA 
 \cite{Yada,Kuroki},
the perturbation theory
 \cite{Nishikawa,Yanase},
and the fluctuation-exchange (FLEX) approximation
 \cite{Mochi,Mochi2}.
When the $d$-orbital freedom is taken into account,
$d$-wave SC due to antiferromagnetic (AF) fluctuations
are dominant when the exchange coupling $J$ is relatively small
 \cite{Yada},
whereas $f(p)$-wave SC due to ferromagnetic (FM) fluctuations
mediated by small hole-pockets occurs for larger $J$
 \cite{Yada,Mochi,Mochi2,Yanase}.
However, only a large $a_{1g}$-like FS is observed by
ARPES measurements by several groups
 \cite{ARPES1,ARPES2}.
Therefore, we have to solve this conflict on the FS's
before studying the mechanism of the SC.

We stress that the electronic properties 
{\it in the normal states} should be analyzed 
in detail before discussing the mechanism of the SC.
In the normal state of Na$_{0.35}$CoO$_2$($\cdot$1.3H$_2$O),
the Knight shift
 \cite{Imai},
the uniform magnetic susceptibility ($\chi^{\rm s}$)
 \cite{Sato-susc}
and the DOS measured by PES
 \cite{Shin}
moderately decrease below 300K.
This ``weak pseudo-gap behaviors'', which is also
observed in high-$T_{\rm c}$ cuprates below
$T_{0}\sim 700$K and above the strong pseudo-gap 
temperature $T^\ast \sim 200$K
 \cite{Sato-Hall,Kontani-Hall},
suggest the occurrence of strong AF-fluctuations
 \cite{Kontani-Hall}.

%%%%%%%%%%%%%%%%%%%%%%%%%%%%%
\begin{figure}[b]
\includegraphics[width=.7\linewidth]{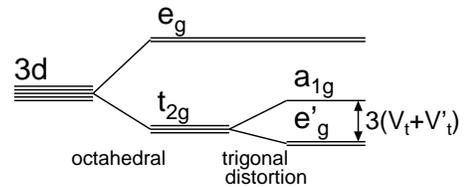}
\caption{CES between $a_{1g}$ and $e_g'$ is caused by
the trigonal deformation of CoO$_6$-octahedron.
$V_t'$ is introduced additionally in the present study.}
\end{figure}
%%%%%%%%%%%%%%%%%%%%%%%%%%%%%

% punch
In the present work, we study the
normal electronic states of Na$_{0.35}$CoO$_2$
to find out the topology of the FS.
Based on the FLEX approximation,
the appropriate weak pseudo-gap behaviors
are well reproduced under the condition that
(i) six small hole-pockets composed of $e_g'$-orbitals
should be absent, and 
(ii) the exchange coupling $J$ should be 
about one order smaller than $U$.
To obtain an appropriate magnitude of the pseudo-gap,
the top of six small hole-pockets, $E_{\rm top}$,
should be just below the chemical potential ($\sim -0.1$eV)
which is consistent with ARPES measurements in 
Na$_{0.35}$CoO$_2$ \cite{ARPES1,ARPES2}.
The pseudo-gap in DOS becomes larger
in water-intercalated samples 
 \cite{Shin},
which would be understood if $E_{\rm top}$
is raised slightly (but still $E_{\rm top}<\mu$),
result of the trigonal deformation of CoO$_6$-octahedron 
by water-intercalation.

% explanation for FLEX
The FLEX approximation is a self-consistent 
spin-fluctuation theory
 \cite{Bickers}.
It can reproduce various non-Fermi-liquid like
behaviors in high-$T_{\rm c}$ cuprates above $T^\ast$
 \cite{Moriya}.
For example, weak pseudo-gap behaviors below $T_0$
in the DOS, $\chi^{\rm s}$ and the specific heat are well reproduced,
are caused by the strong AF-fluctuations.
Here, we analyze the role of the
spin fluctuations on the electronic states
of Na$_x$CoO$_2$, using the FLEX approximation.
We note that the FLEX approximation for 
systems with orbital degeneracy is so complicated that
only a few works have been performed previously
 \cite{Mochi,Mochi2}.

Here we explain in detail the $d$-$p$ model Hamiltonian
and the formalism of the FLEX approximation
with orbital degree of freedom.
The unit of energy is eV hereafter.
The kinetic term of the Hamiltonian is
\begin{eqnarray}
H_0&=&\sum_{i,j,\sigma}\sum_{\ell\ell'}t^{\ell\ell'}_{ij} \ c^\dag_{i\ell\sigma} c_{j\ell'\sigma} \nonumber \\
   &=&\sum_{{\bf k},\sigma}\sum_{\ell\ell'}\varepsilon_{\bf k}^{\ell\ell'} \ c^\dag_{{\bf k}\ell\sigma} c_{{\bf k}\ell'\sigma} ,
 \label{eqn:H0}
%\\
%&=&\sum_{{\bf k},\alpha,\sigma}\tilde\varepsilon_{{\bf k}\alpha} \ a^\dag_{{\bf k}\alpha\sigma} a_{{\bf k}\alpha\sigma}
\end{eqnarray}
where
\begin{eqnarray*}
c_\ell=\left\{
\begin{array}{ccc}
d_{xy},d_{yz},d_{zx},d_{x^2-y^2},d_{3z^2-r^2} & : & \mbox{Co} \\
p_{1x},p_{1y},p_{1z} & : & \mbox{O(1)} \\
p_{2x},p_{2y},p_{2z} & : & \mbox{O(2)} \\
\end{array}
\right.
\end{eqnarray*}
Here, $t^{\ell\ell'}_{ij}$ in eq. (\ref{eqn:H0})
is given by the following Slater-Koster parameters:
$(pd\pi)=0.78$, $(pd\s)=-1.7$, $(pp\pi)=0.18$ and
$(pp\s)=-0.48$ for nearest-neighbor transfers,
$(pp\pi)_2=0.11$ and $(pp\s)_2=-0.26$
for next-nearest ones.
$10Dq=1.1$ is the CES between $t_{2g}$ and $e_g$,
and $\Delta=1.8$ is the $d$-$p$ charge transfer energy.
Reflecting the trigonal deformation, furthermore,
$3V_t=-0.18$ is the CES between $a_{1g}$ and $e_g'$,
$c_t=1.3$ is the ratio of the inter-layer $pp$ transfer
to the intra-layer one, and $(pd)_t=0.31$ is the
$pd$ transfer emerged by the deviation of O-Co-O angle from $\pi$. 
They are determined by fitting the Singh's LDA-band structure
 \cite{Yada}.

%%%%%%%%%%%%%%%%%%%%%%%%%%%%%
\begin{figure}
\includegraphics[width=.7\linewidth]{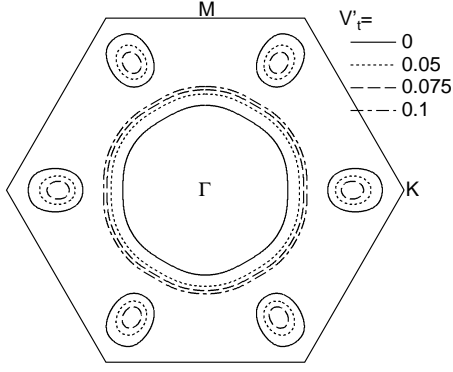}% Here is how to import EPS art
\caption{Fermi surfaces for several values of $V_t'$, 
where the density of hole is 0.65.
Small FS's disappear when $V_t'\ge0.1$.
A lerge FS around the $\Gamma$-point and six small FS's
are mainly composed of $a_{1g}$-orbital and $e_g'$-ones, respectively.}
\end{figure}
%%%%%%%%%%%%%%%%%%%%%%%%%%%%%

Here we introduce $V_t'$ to express the change of the CES.
The FS's given by the FLEX approximation are shown
in Figs. 2, where the density of hole is fixed at 0.65.
$V_t'=0$ corresponds to the original Singh's FS's.
The shape of FS's is sensitive to $V_t'$.
The small hole-pockets disappear when $V_t'\ge0.1$, 
whose band-structure is similar to the experimental 
results by ARPES measurements
 \cite{ARPES1,ARPES2}.
We find that the deformation of the FS's 
due to the interaction is tiny in the present model,
which is consistent with
the analysis by the Gutzwiller approximation
 \cite{Sugi}.

In addition to $H_0$,
we treat the on-site correlation in $t_{2g}$-orbitals,
which is expressed in a symmetric form as
\begin{eqnarray}
H'
%&=&\frac{1}{2}\sum_i\sum_{\zeta_1\zeta_2\zeta_3\zeta_4}I_{\zeta_1\zeta_4,\zeta_3\zeta_2}^{(0)}
%c^\dag_{i\zeta_1}c^\dag_{i\zeta_2}c_{i\zeta_3}c_{i\zeta_4}\\
&=&\frac{1}{4}\sum_i\sum_{\zeta_1\zeta_2\zeta_3\zeta_4}\Gamma_{\zeta_1\zeta_4,\zeta_3\zeta_2}^{(0)}
c^\dag_{i\zeta_1}c^\dag_{i\zeta_2}c_{i\zeta_3}c_{i\zeta_4} ,
 \label{eqn:Hd2}
\end{eqnarray}
where $\zeta\equiv (l,\sigma)$.  
${\hat \Gamma}^{(0)}$ is given by
\begin{eqnarray}
& &\Gamma^{(0)}_{\zeta_1\zeta_4,\zeta_3\zeta_2}=-\frac{1}{2}S^{(0)}_{\ell_1\ell_4,\ell_3\ell_2}\mbox{\boldmath$\sigma$}_{\sigma_1\sigma_4}\cdot\mbox{\boldmath$\sigma$}_{\sigma_2\sigma_3}
 \nonumber \\
& & \hspace{5.3em}
 +\frac{1}{2}C^{(0)}_{\ell_1\ell_4,\ell_3\ell_2}\delta_{\sigma_1\sigma_4}\delta_{\sigma_2\sigma_3}
 \label{eqn:G}, \\
& &\hat S^{(0)}=\left\{
\begin{array}{c}
U\\
U'\\
J\\
J'\\
\end{array}
\right. \hspace{0em},
\hat C^{(0)}=\left\{
\begin{array}{cc}
\hspace{-0.5em}U&\hspace{-1em}\ \ (\ell_1=\ell_2=\ell_3=\ell_4)\\
\hspace{-0.5em}-U'+2J&\hspace{-1em}\ \ (\ell_1=\ell_3\ne\ell_2=\ell_4)\\
\hspace{-0.5em}2U'-J&\hspace{-1em}\ \ (\ell_1=\ell_4\ne\ell_2=\ell_3)\\
\hspace{-0.5em}J'&\hspace{-1em}\ \ (\ell_1=\ell_2\ne\ell_3=\ell_4)\\
\end{array}
\right.\nonumber
%
%S^{(0)}_{\ell_1\ell_4,\ell_3\ell_2}
%&=& ....
% \\
%C^{(0)}_{\ell_1\ell_4,\ell_3\ell_2}
%&=& ....
\end{eqnarray}
where $U$ ($U'$) is the intra-orbital (inter-orbital)
direct Coulomb interaction,
and $J$, $J'$ represent the exchange interactions.
We put $J=J'$ hereafter.

In the FLEX approximation, 
the spin and charge susceptibility are given by
\begin{eqnarray}
\hat \chi^s(q)&=&\left({\bf 1}-\hat S^{(0)}\hat\chi^{(0)}(q)\right)^{-1}\hat \chi^{(0)}(q) ,
 \label{eqn:chis} \\
\hat \chi^c(q)&=&\left({\bf 1}+\hat C^{(0)}\hat\chi^{(0)}(q)\right)^{-1}\hat \chi^{(0)}(q) ,
  \label{eqn:chic}
\end{eqnarray}
where $q=(\q,\w_l=2\pi T l)$.
$\hat \chi^{(0)}(q)$ is given by
\begin{eqnarray}
\chi^{(0)}_{\ell_1\ell_2,\ell_3\ell_4}(q)=-\frac{T}{N}\sum_kG_{\ell_1\ell_3}(k+q)G_{\ell_4\ell_2}(k) ,
\end{eqnarray}
where $k=(\k,\e_n=\pi T(2n+1))$. 
$\hat G(k)=\left( \{ \hat G^{(0)}(k) \}^{-1}-
 \hat\Sigma(k)\right)^{-1}$, and
$\hat G^{(0)}(k)=\left(\w+\mu-{\hat \e}_\k\right)^{-1}$.
The self-energy is given by
\begin{eqnarray}
\Sigma_{\ell_1\ell_2}(k)
 &=& \frac{T}{N}\sum_q\sum_{\ell_3\ell_4}G_{\ell_3\ell_4}(k-q)
 V^{\rm eff}_{\ell_1\ell_3,\ell_2\ell_4}(q) ,
 \label{eqn:Sigma}
\end{eqnarray}
where $\hat V^{\rm eff}(q)= \frac{3}{2}\hat V^s(q)+\frac{1}{2}\hat V^c(q)$;
\begin{eqnarray*}
\hat V^s(q)&=& {\hat S^{(0)}} {\hat \chi^s(q)} {\hat S^{(0)}}
 -\frac{1}{2}\hat S^{(0)}\hat\chi^{(0)}(q)\hat S^{(0)}  ,\\
\hat V^c(q)&=&  {\hat C^{(0)}} {\hat \chi^c(q)} {\hat C^{(0)}}
 -\frac{1}{2}\hat C^{(0)}\hat\chi^{(0)}(q)\hat C^{(0)} .
\end{eqnarray*}
We solve eqs. (\ref{eqn:chis})-(\ref{eqn:Sigma}) self-consistently.
In the numerical study, we use 4096 $\k$-meshes and
256 Matsubara frequencies.

%%%%%%%%%%%%%%%%%%%%%%%%%%%%%%%%%%
\begin{figure*}
\includegraphics[width=.8\linewidth]{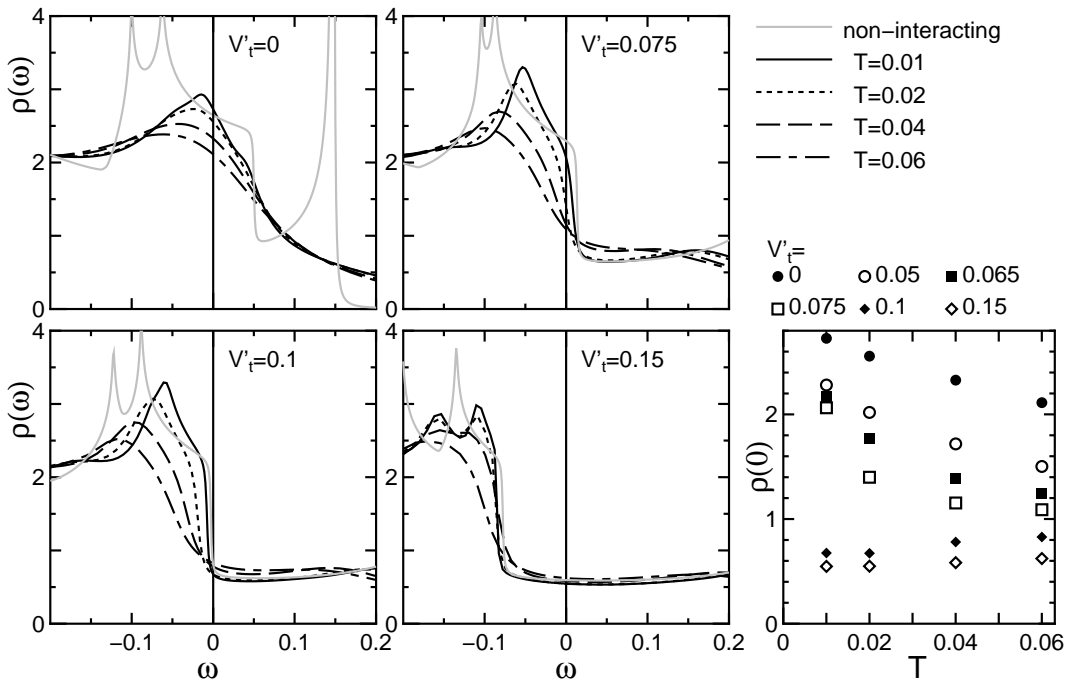}% Here is how to import EPS art
\caption{Obtained $\rho(\w)$ for several values of $V_t'$.
When $V_t'\le 0.075$ ($V_t'\ge 0.1$),
$\rho(0)$ decreases (increases) as $T \rightarrow0$.
$T=0.01$[eV] corresponds to 120K.}
\end{figure*}
%%%%%%%%%%%%%%%%%%%%%%%%%%%%%%%%%%

Hereafter,
we present the results by the FLEX approximation for 
$0\le V_t'\le 0.15$.
We will see that the appropriate
weak pseudo-gap behaviors are reproduced only for $V_t'\ge 0.1$,
where a large single FS around the $\Gamma$-point exists.
%This results means that the value of $V_t$ 
%obtained by fitting the Singh's LDA study would be incorrect.
Note that $V_t'$ will ``decrease'' by water-intercalation 
because trigonal deformation, which raise the $e_g'$-level,
is promoted further
 \cite{structure}.
In the present study, we put $U'=1.3$, $J=0.13$ 
($U=U'+2J=1.56$), where AF-fluctuations dominate the 
FM-ones as shown below.
We have checked that similar results are obtained
if we put $J= 0.26$.
On the other hand, ref. \cite{Mochi2}
reports that strong FM-fluctuations occur
when $J/U=1/5$ (i.e., $J/U'=1/3$).
These results are consistent with the magnetic
phase diagram in the present model obtained by
the RPA \cite{Yada}; $U$ promotes the AF-fluctuations
whereas $J$ promotes the FM-ones.

Figure 3 shows the
temperature dependences of the DOS,
$\rho(\w)=\sum_\k {\rm Tr \{ Im}{\hat G}_\k(\w-i\delta) \} /\pi$.
$\w=0$ corresponds to the chemical potential.
First, we discuss the case of $V_t'\le0.075$;
then, the DOS without interaction, $\rho^0(\w)$,
is very large at $\w=0$ due to six small hole-pockets.
In the presence of interaction,
$\rho(0)$ at higher temperatures is small
because huge Im${\hat \Sigma}_\k(0)$ by the
FLEX approximation smears the steep $\w$-dependence 
of $\rho^0(\w)$.
As the temperature decreases,
the DOS obtained by the FLEX approximation
approaches $\rho^0(\w)$ because the quasiparticles
become coherent.
As a result,  $\rho(0)$ increases as the temperature
decreases, which may be called the ``anti-pseudo-gap behavior''.

In case of $V_t'\ge0.1$, on the other hand,
$\rho^0(0)$ is small because six small hole-pockets 
sink below $\mu$.
Therefore, Im${\hat \Sigma}_\k(0)$ by the FLEX approximation
is relatively small.
Then, $\rho(0)$ in the presence of interaction 
slightly decreases as $T\rightarrow 0$, reflecting
the reduction of the DOS around the ``hot-spot'' on the FS, 
where Im${\hat \Sigma}_\k(0)$ takes the maximum value
due to strong AF-fluctuations.
%When AF-fluctuations are strong, Im${\hat \Sigma}_\k(0)$ 
%depends on $\k$ prominently, and the quasiparticles around 
%the hot-spot becomes incoherent.
Each hot-spot is connected with others by the nesting vector.
We find that the hot-spot lays on the cross-point of
a large FS and a $\Gamma$-K line for $V_t'\ge0.1$.
As a result,  $\rho(0)$ slightly decreases as the temperature
decreases, which we call the ``weak pseudo-gap behavior''.
We note that the FLEX approximation tends to 
underestimates the size of the pseudo-gap;
the vertex corrections for $\Sigma_\k(\w)$
might recover its correct size
 \cite{Pines-PG}.

%%%%%%%%%%%%%%%%%%%%%%%%%%%%%%%%%%
\begin{figure*}[t]
\includegraphics[width=.8\linewidth]{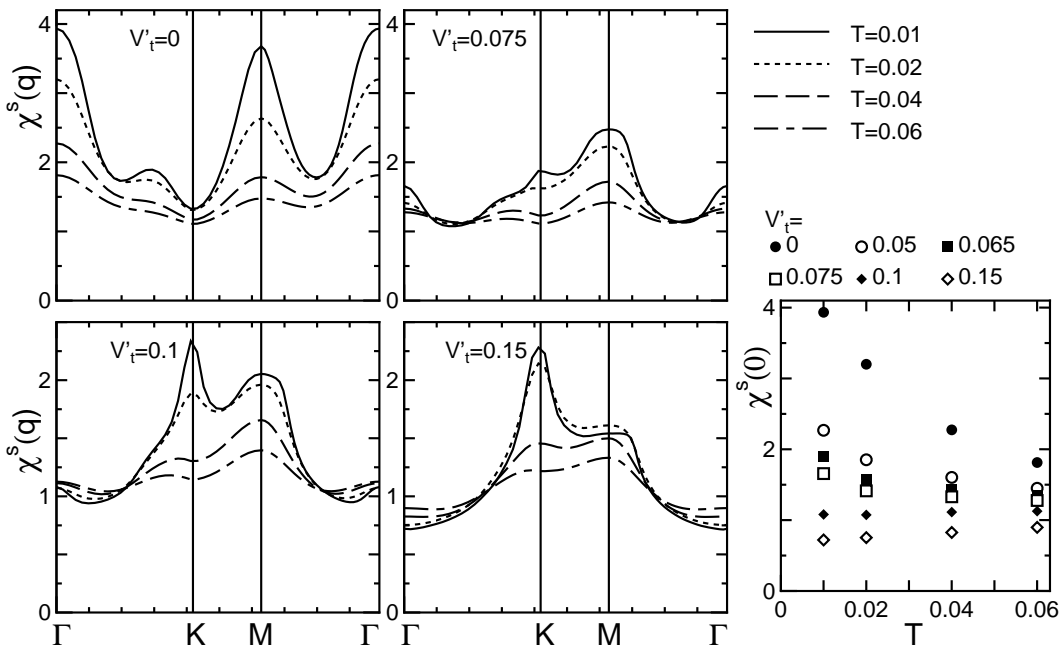}% Here is how to import EPS art
\caption{Obtained $\chi^{\rm s}({\bf q},0)$ for several values of $V_t'$.
When $V_t'\le 0.075$ ($V_t'\ge 0.1$),
the uniform susceptibility decreases (increases) as $T \rightarrow0$,
and $\chi^{\rm s}({\bf q},0)$ takes the maximum value at 
$\q=(0,2\pi/\sqrt{3})$ ($\q=(4\pi/3,0)$).
}
\end{figure*}
%%%%%%%%%%%%%%%%%%%%%%%%%%%%%%%%%%

Figure 4 shows the
temperate dependences of the spin susceptibility
$\chi^{s}(\q,0)= {\rm Tr} {\hat \chi}^{s}(\q,0)$
for several values of $V_t'$.
When small hole-pockets exist ($V_t'\le0.075$),
both the FM-fluctuations (at $\q=0$) as well as the
AF-fluctuations with $\q={\vec{\Gamma M}}=(0,2\pi/\sqrt{3})$
grow as the temperature decreases.
The latter come from the nesting of small hole-pockets
between neighboring Brillouin zones,
whereas the former is caused by each small hole-pockets.
Thus, small hole-pockets govern the magnetic fluctuations
when $V_t'\le0.075$.
On the other hand, when small hole-pockets are absent 
($V_t'\ge0.1$), AF-fluctuations with 
$\q={\vec{\Gamma K}}=(4\pi/3,0)$,
which come from the nesting of a large FS around the 
$\Gamma$-point, dominate the fluctuations with $\q=(0,2\pi/\sqrt{3})$.
On the contrary, 
the uniform susceptibility ($\q=0$) slightly decreases
as $T\rightarrow0$, reflecting the emergence of the
pseudo-gap in the DOS as shown above.

As a result, experimental weak pseudo-gap behaviors
are qualitatively reproduced only when the
small hole-pockets are absent ($V_t'\ge0.1$).
Because the pseudo-gap phenomena are caused by 
strong AF-fluctuations, they cannot be reproduced
when $J/U$ is so large that FM-fluctuations dominates.
Thus, we conclude that (i)small hole-pockets are absent,
and (ii) $J/U\sim O(1/10)$.

Here we discuss the effect of water-intercalation.
Experimentally, both for Na$_{0.35}$CoO$_2$ and
Na$_{0.35}$CoO$_2 \cdot$1.3H$_2$O, 
the uniform susceptibility 
 \cite{Sato-susc} as well as the Knight shift \cite{Imai}
slightly decrease below room temperatures,
if impurity effect is subtracted.
In water-intercalated samples,
the size of the pseudo-gap in DOS is larger
 \cite{Shin}, and prominent Curie-Weiss behavior of $1/T_1T$ 
is observed below 100K, which
suggests the emergence of strong AF-fluctuations
 \cite{Sato-susc}.
Comparing these experimental facts with
the theoretical results, we infer that 
Na$_{0.35}$CoO$_2$ and Na$_{0.35}$CoO$_2 \cdot$1.3H$_2$O
correspond to $V_t'=0.15$ and 0.1, respectively.
Note that $V_t'$ will be reduced by the water-intercalation,
as discussed before.
However, small hole-pockets are still absent ($V_t'\ge0.1$)
even in Na$_{0.35}$CoO$_2 \cdot$1.3H$_2$O, because Knight shift,
the specific heat, and $1/T_1T$ above 100K are almost
unchanged by the water-intercalation.

%%%%%%%%%%%%%%%%%%%%%
% Summary
%%%%%%%%%%%%%%%%%%%%%
In summary, we analyzed the $d$-$p$ Hubbard model for 
Na$_{0.35}$CoO$_2$ using the FLEX approximation, and
deduced that {\it small hole-pockets are absent}:
When $J$ is relatively small ($J\sim U/10$),
strong AF-fluctuations due to the nesting emerge.
Thanks to the AF-fluctuations, 
experimental weak pseudo-gap behaviors below room temperature 
are well reproduced, only when top of six small hole-pockets 
is just below the Fermi level.
Then, the obtained mass-enhancement ratio ($\sim 2$) 
%as well as Wilson ratio ($\sim 2-3$)
is consistent with the renormalization of bandwith
observed by ARPES
 \cite{ARPES1,ARPES2}
as well as the analysis by the Gutzwiller approximation
 \cite{Sugi}.
On the contrary, weak pseudo-gap does not emerge for larger $J$
where FM-fluctuations are dominant, which is inconsistent
with experiments.
In later works, the mechanism of the SC should be studied
based on the model established in the present work,
with only a single FS composed of $a_{1g}$-orbital.
We will also perform the study below 100K in future.

%%%%%%%%%%%%%%%%%%%%%
\acknowledgements
We are grateful for M. Sato and Y. Kobayashi 
for valuable discussions on experiments,
and for D.S. Hirashima and K. Yamada
for useful comments and discussions.

%%%%%%%%%%%%%%%%%%%%
% references
%%%%%%%%%%%%%%%%%%%%

\end{document}